# Concept Combination and the Origins of Complex Cognition

**Liane Gabora**

liane.gabora@ubc.ca

University of British Columbia

Okanagan campus, Department of Psychology, Arts Building

3333 University Way

Kelowna BC, V1V 1V7, CANADA

**Kirsty Kitto**

kirsty.kitto@qut.edu.au

Queensland University of Technology

Information Systems School

Brisbane, AUSTRALIA

ABSTRACT

At the core of our uniquely human cognitive abilities is the capacity to see things from different perspectives, or to place them in a new context. We propose that this was made possible by two cognitive transitions. First, the large brain of *Homo erectus* facilitated the onset of *recursive recall:* the ability to string thoughts together into a stream of potentially abstract or imaginative thought. This hypothesis is supported by a set of computational models where an artificial society of agents evolved to generate more diverse and valuable cultural outputs under conditions of recursive recall. We propose that the capacity to see things in context arose much later, following the appearance of anatomically modern humans. This second transition was brought on by the onset of *contextual focus*: the capacity to shift between a minimally contextual analytic mode of thought, and a highly contextual associative mode of thought, conducive to combining concepts in new ways and 'breaking out of a rut'. When contextual focus is implemented in an art-generating computer program, the resulting artworks are seen as more creative and appealing. We summarize how both transitions can be modeled using a theory of concepts which highlights the manner in which different contexts can lead to modern humans attributing very different meanings to the interpretation of one concept.

What is the essence of our human-ness? We propose that what is at the core of our uniquely human cognitive abilities is the capacity to *place things in context*, or *see things from different perspectives*. This enables us to be not just creative, but to put our own spin on the inventions of others, modifying them to suit our own needs and tastes, in turn leading to new innovations that build cumulatively on previous ones (Gabora, 2003, 2008; Gabora & Russon, 2011; Gabora & Sternberg, 2010). It enables us to modify thoughts, impressions, and attitudes by thinking about them in the context of each other, and thereby weave them into a more or less integrated structure that defines who we are in relation to the world. Our compunction to put our own spin on the ideas and inventions of others results in accumulative cultural change, referred to as the *ratchet effect* (Tomasello, Kruger, & Ratner, 1993).

Understanding how this capacity evolved, and testing it against other theories about what is responsible for our human-ness, is difficult. All that is left of our prehistoric ancestors are their bones and artifacts such as stone tools that resist the passage of time. Methods for analyzing these remains are becoming increasingly sophisticated, but they still leave many questions unanswered, and are often compatible with several competing theories. Thus, in seeking to explain the evolution of the uniquely human cognitive capacities that have transformed our lives, and even the planet we live on, formal computational and mathematical models provide an extremely valuable set of reconstructive tools. Steps toward a mathematical model of the evolution of the cognitive mechanisms underlying the evolution of the capacity to 'see things in context' has been put forward (Gabora & Aerts, 2009), and computational models of this have also been developed (DiPaola & Gabora, 2007, 2009; Gabora, 1994, 1995, 2008a,b; Gabora & Leijnen, 2009; Leijnen & Gabora, 2009, 2010; Gabora, Leijnen, & von Ghyczy, in press; Gabora & Saberi, 2011). The goal of this chapter is to explain these efforts in layperson terms, which fill in some gaps, and show how they constitute an integrated effort to formally model the evolution of the cognitive mechanisms that underlie our human-ness.

**First Transition: The Earliest Signs of Creativity**

The last common ancestor of humans and other great apes lived between four and eight million years ago. The minds of our earliest ancestors, *Homo habilis,* have been referred to as *episodic* because there is no evidence that their experience deviated from the present moment of concrete sensory perceptions (Donald, 1991). They were able to encode perceptions of events in memory, and recall them in the presence of a reminder or cue, but had little voluntary access to memories without environmental cues. They would for example not think of a particular person or object unless something in their environment concretely triggered its recall. They were therefore unable to voluntarily shape, modify, or practice skills and actions, and neither could they invent or refine complex gestures or means of communicating.

*Homo habilis* was eventually replaced by *Homo erectus,* which lived between approximately 1.8 and 0.3 million years ago. This period is widely referred to as the beginnings of human culture. The cranial capacity of the *Homo erectus* brain was around 1,000 cc, which is about 25% larger than that of *Homo habilis*, at least twice as large as that of living great apes, and 75% that of modern humans (Aiello, 1996; Ruff et al., 1997). *Homo erectus* exhibited many indications of enhanced intelligence, creativity, and an ability to adapt to their environment. For example, they made use of sophisticated task-specific stone hand axes, complex stable seasonal home bases, and long-distance hunting strategies involving large game, and migration out of Africa.

This period marks the onset of the archaeological record and it is thought to be the beginnings of human culture. It is widely believed that this cultural transition reflects an underlying transition in cognitive or social abilities. Some have suggested that such abilities arose with the onset of a *theory of mind* (Mithen, 1998) or the capacity to imitate (Dugatkin, 2001). However, there is evidence that nonhuman primates also possess theory of mind and the capacity to imitate (Heyes, 1998; Premack, 1988; Premack, & Woodruff, 1978), and yet they do not compare to modern humans in intelligence and cultural complexity.

Evolutionary psychologists have suggested that the intelligence and cultural complexity of the *Homo* line is due to the onset of *massive modularity* (Buss, 1999, 2004; Barkow, Cosmides, & Tooby, 1992). However, although the mind exhibits an intermediate degree of functional and anatomical modularity, neuroscience has not revealed vast numbers of hardwired, encapsulated, task-specific modules; indeed, the brain has been shown to be more highly subject to environmental influence than was previously believed (Buller, 2005; Byrne, 2000; Wexler, 2006).

**A Promising and Testable Hypothesis**

Donald (1991) proposed that with the enlarged cranial capacity of *Homo erectus*, the human mind underwent the first of three transitions by which it—along with the cultural matrix in which it is embedded—evolved from the ancestral, pre-human condition. This transition is characterized by a shift from an *episodic* to a *mimetic mode* of cognitive functioning, made possible by onset of the capacity for voluntary retrieval of stored memories, independent of environmental cues. Donald refers to this as a *self-* *triggered recall and rehearsal loop*. Self-triggered recall enabled information to be processed recursively, and reprocessed with respect to different contexts or perspectives. This allowed our ancestors to access memories voluntarily and thereby to act out[1] events that occurred in the past or that might occur in the future. Thus not only could the mimetic mind temporarily escape the here and now, but by miming or gesture it could communicate similar escapes to other minds. The capacity to mime thus brought forth what is referred to as a *mimetic* form of cognition, so ushering in a transition to the mimetic stage of human culture. The self-triggered recall and rehearsal loop also enabled our ancestors to engage in a stream of thought, where one thought or idea evokes another, revised version of it, which evokes yet another, and so forth recursively. In this way, attention is directed away from the external world toward one's internal model of it. Finally, self-triggered recall allowed for voluntary rehearsal and refinement of actions, enabling systematic evaluation and improvement of skills and motor acts.

**Computational Model**

The recursive recall hypothesis is difficult to test directly, for even if correct, the brain tissues of our ancestors are long disintegrated, so we cannot directly study how the neural mechanisms underlying recursive recall evolved. It is, however, possible to computationally model how the onset of the capacity for recursive recall would affect the effectiveness, diversity, and open-endedness of ideas generated in an artificial society. This section summarizes how we tested Donald's hypothesis using an agent-based computational model of culture referred to as 'EVOlution of Culture', abbreviated EVOC. Details of the modeling platform are provided elsewhere (Gabora, 2008b, 2008c; Gabora & Leijnen, 2009; Leijnen & Gabora, 2009).

**The EVOC World**. EVOC uses neural network based agents that (i) invent new ideas, (ii) imitate actions implemented by neighbors, (iii) evaluate ideas, and (iv) implement successful ideas as actions. Invention works by modifying a previously learned action using learned trends (such as that more overall movement tends to be good) to bias the invention process. The process of finding a neighbor to imitate works through a form of lazy (what computer scientists refer to as 'non-greedy', by which they mean that solutions provided at each stage in an iteration are not necessarily optimal) search. An imitating agent randomly scans its neighbors, and assesses the fitness of their actions using a predefined fitness function. It adopts the first action that is fitter than the action it is currently implementing. If it does not find a neighbor that is executing a fitter action than its own action (see below for a discussion of fit-ness), it continues to execute the current action. Over successive rounds of invention and imitation, agents' actions improve. EVOC thus models how descent with modification occurs in a purely cultural context. Agents do not evolve in a biological sense–they neither die nor have offspring–but do in a cultural sense, by generating and sharing ideas for future actions.[2]

[1] The term *mimetic* is derived from "mime," which means "to act out."

Following Holland (1975) we refer to the success of an action in the artificial world as its *fitness,* with the caveat that unlike its usage in biology, here the term is unrelated to number of offspring (or ideas derived from a given idea). The fitness function rewards head immobility and symmetrical limb movement. Fitness of actions starts out low because initially all agents are entirely immobile. However, some agent quickly invents an action that has a higher fitness than doing nothing, and this action gets imitated, leading to an increase in fitness. Fitness increases further as other ideas get invented, assessed, implemented as actions, and spread through imitation. The diversity of actions initially increases due to the proliferation of new ideas, and then decreases as agents hone in on the fittest actions.

The artificial society consists of a toroidal lattice with 100 nodes, each occupied by a single, stationary agent. We used a von Neumann neighborhood structure (agents only interacted with their four adjacent neighbors). During invention, the probability of changing the position of any body part involved in an action was 1/6. On each run, creators and imitators were randomly dispersed.

**Chaining.** This gives agents the opportunity to execute multi-step actions. For the experiments reported here with chaining turned on, if in the first step of an action an agent was moving at least one of its arms, it executes a second step, which again involves up to six body parts. If, in the first step, the agent moved one arm in one direction, and in the second step it moved the same arm in the opposite direction, it has the opportunity to execute a three-step action. And so on. The agent is allowed to execute an arbitrarily long action so long as it continues to move the same arm in the direction opposite to the direction it moved previously. Once it does not do so, the chained action comes to an end. The longer it moves, the higher the fitness of this multi-step chained action. This is admittedly a simple action, but we were not interested in the impact of this action *per se*. The goal here was simply to test hypotheses about how chaining at the individual level affects dynamics at the societal level, by providing agents with a means of implementing multistep actions such that the optimal way of going about one step depends on how one went about the previous step. This seems to be a common feature of many useful actions such as the repetitive motions involved in toolmaking, sawing, carving, weaving, and so forth.

The fitness of a chained action was calculated as follows. Where *c* is 'with chaining', *w* is 'without

[2] For an explanation of why we do not adopt the framework of memetics see (Gabora, 1999, 2004, 2008d).

chaining', $n$ is the number of chained actions, the fitness, $F_c$, is calculated as follows:

$$F_c = F_w(n-1)$$

The fitness function with chaining provides a simple means of simulating the capacity for recursive recall.

## Results

As shown in Figure 1, the capacity to chain together simple actions to form more complex ones increases the mean fitness of actions across the artificial society. This is most evident in the later phase of a run. Without chaining, agents converge on optimal actions, and the mean fitness of action reaches a plateau. With chaining, however, there is no ceiling on the mean fitness of actions. By the 200$^{th}$ iteration the chaining process has led to more than double the maximum fitness attainable without chaining.

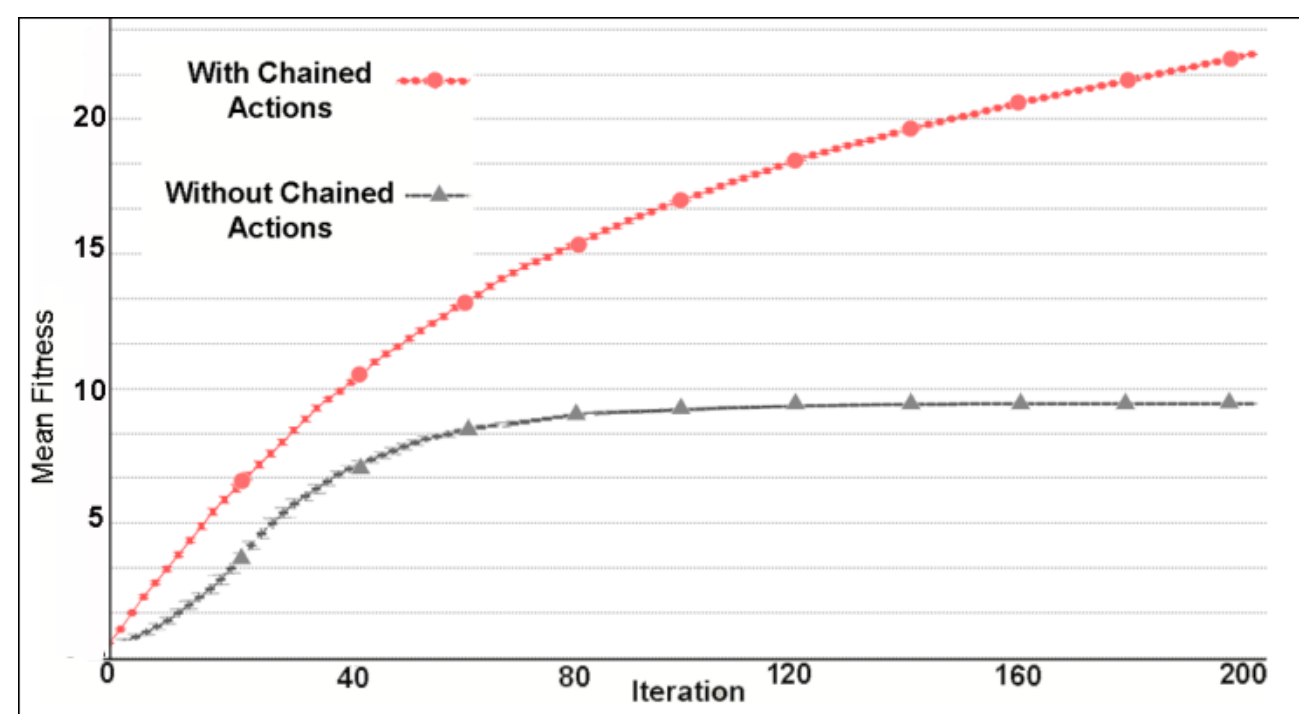

Figure 1. Mean fitness of actions in the artificial society with chaining versus without chaining. From (Gabora & Saberi, 2011).

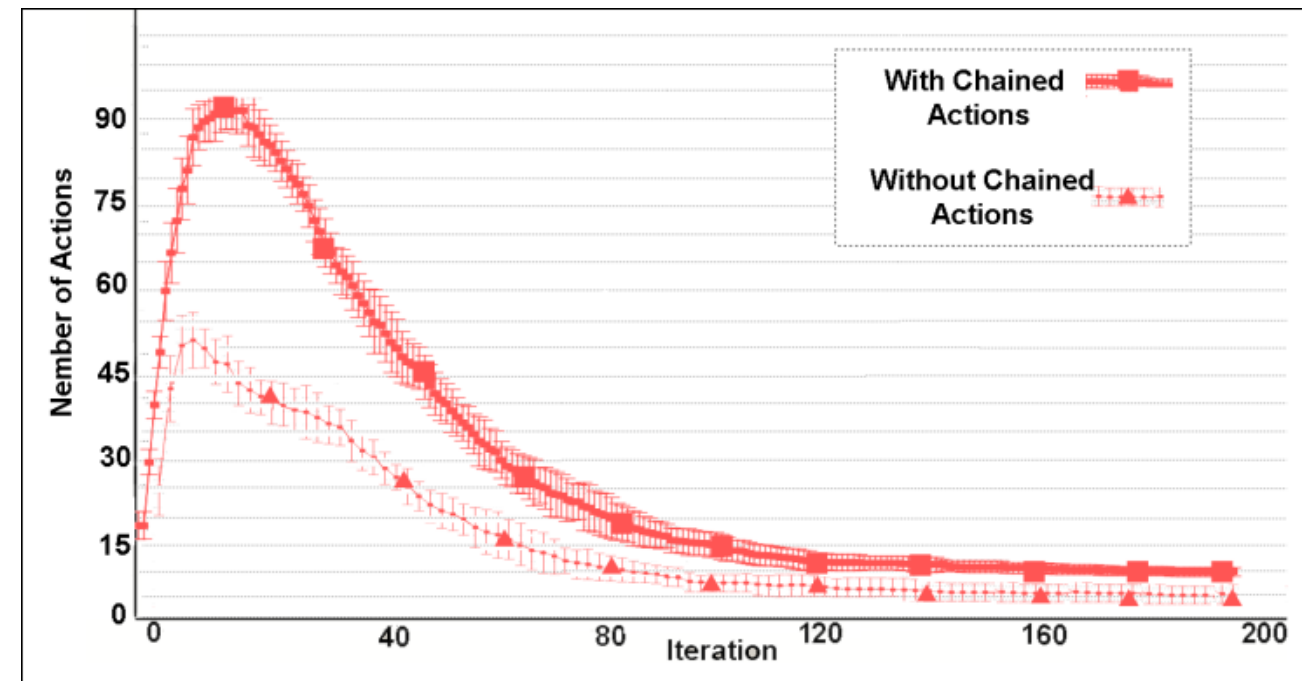

Figure 2. Mean number of different actions in the artificial society with chaining (continuous line) versus without chaining (dashed line). From (Gabora & Saberi, 2011).

As shown in Figure 2, chaining also increases the diversity of actions. This is most evident in the early phase of a run before agents begin to converge on optimal actions. Although in both cases there is convergence on optimal actions, without chained actions, this is a static set (thus mean fitness plateaus) whereas with chained actions the set of optimal actions is always changing, as increasingly fit actions are found (thus mean fitness keeps increasing).

This shows that recursive recall increased the fitness of ideas while simultaneously increasing the number of different ideas across the artificial society. It thus supports the hypothesis that the onset of recursive recall was a critical step toward the kind of cognition we associate with humans.

We also tested the effect of chaining on the capacity to benefit from learning. Recall that agents have the capacity to learn trends from past experiences, and thereby bias the generation of novelty in directions that have a greater than chance probability of being fruitful. Since chaining provides more opportunities to capitalize on the capacity to learn, we hypothesized that chaining would accentuate the impact of learning on the mean fitness of actions, and this too turned out to be the case (Gabora & Saberi, 2011).

Note that in the chaining versus no chaining conditions the size of the neural network is the same, but how it is used differs. This suggests that it was not larger brain size per se that initiated the onset of cumulative culture, but that larger brain size enabled episodes to be encoded in more detail, allowing more routes for reminding and recall, thereby facilitating the ability to recursively re-describe information stored in memory (Karmiloff-Smith, 1992), and thereby to tailor it to the situation at hand.

**Mathematical Modeling of Recursive Re-description: An Idea in Context**

A limitation of this model is that the recursive recall does not work, as it does in humans, by considering an idea in light of one perspective, seeing how that perspective modifies the idea, recognizing in what respect this modification suggests a new perspective from which to consider the idea, and so on. Mathematical modeling of recursive re-description requires an approach that can incorporate the effect of context on the state of a context. It is widely recognized that the standard analytical techniques of science are not up to the challenge of modeling these contextual effects. When concepts appear in the context of each other, their meanings change in ways that are non-compositional, i.e., they behave in ways that violate the rules of classical logic (Osherson & Smith, 1981; Hampton, 1987; Aerts & Gabora, 2005; Kitto, 2006, 2008a; Aerts, 2009; Kitto, Ramm, Sitbon, & Bruza, 2011). Despite its potential im-pact, this challenge is not as insurmountable as it might at first seem, as there is one mathematical formalism which was invented precisely to describe such contextuality; Quantum Theory (QT). It is not the purpose of this paper to describe either this theory,[3] nor applications of it to cognition, in any detail. Rather, we seek here to motivate the model discussed in Kitto *et. al* (2012) as a viable formalism that can describe the density of information storage that is required before the transition to recursive redescription can take place.

This model explicitly represents the context in which information occurs via a notion of measurement. Put simply, for quantum systems, a measurement does not simply record what is there, but interacts with the system under consideration to reveal information about its state *in the context defined by the measurement setting* (Kitto, 2008b). In this theory, it is impossible to refer to the state of a system without reference to a measurement setting. Similarly, considering some concept *w* without reference to the context in which it occurs is implausible at best. FIRE for example might be a danger (in a FOREST FIRE), a tool (a COOKING FIRE), a light source and community hub (a CAMP FIRE) etc., and the meaning that we attribute to the concept FIRE will vary widely as a result. In Bruza *et. al,* (2009) a simple model of this effect as it applies to the human mental lexicon was presented, and here we shall briefly overview that model. In particular, we shall illustrate the manner in which the same idea can be attributed with more than one meaning, so contributing to the density of information storage without requiring an associated increase in physical storage (*i.e.*, neural capacity).

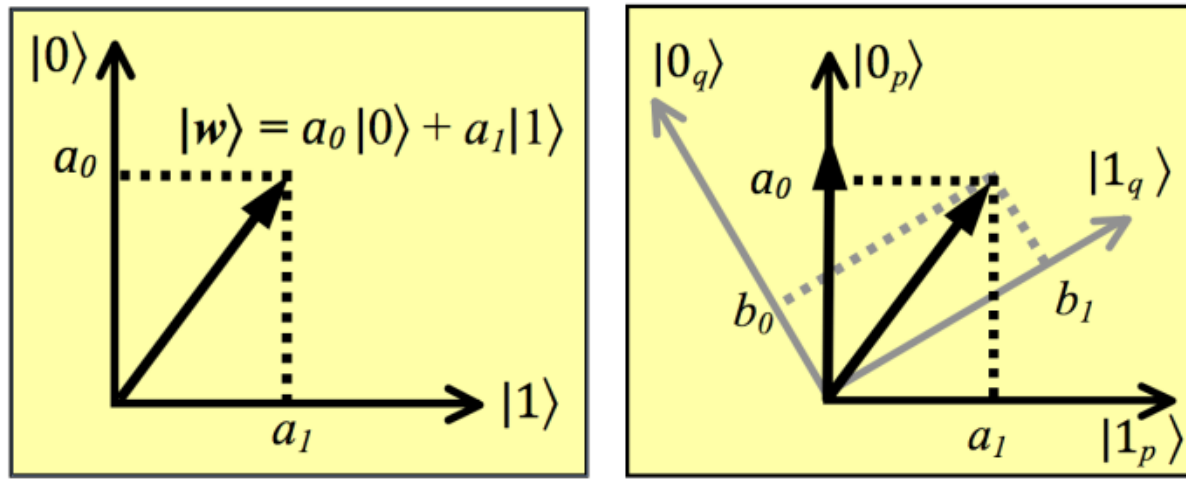

Figure 3. Representing the idea |*w*› in two different contexts. (a) In the context represented by *p*, an idea has a probability of being interpreted with a particular meaning that is related to the projection of the idea onto the context represented by a set of basis states. (b) In a different context *q*, this probability changes markedly, which can be seen by the different projections onto the new context.

[3] This is summarized nicely in Isham, (1995).

In Figure 3, we have drawn a *geometrical* representation of an idea in context. Here, the idea $|w\rangle$ is drawn as the same vector, but in two different contexts, representing its potential to be interpreted $|1\rangle$, or not $|0\rangle$, in the sense represented by that context. For example, a concept of FIRE represented in a particular context will have a certain potential of being interpreted as *dangerous* by a person (e.g., FIRE is almost always interpreted as dangerous by the residents of Australia during summer.) Thus, unless the vector is perfectly aligned with one of the axes in the diagram, then a person who is asked about that concept will be genuinely undecided as to how they will interpret it. We represent this genuine indecision as a *superposition state*:

$$|w\rangle = a_0|0_p\rangle + a_1|1_p\rangle, \text{ where } |a_0|^2 + |a_1|^2 = 1.$$

However, in the different context represented by Fig. 3(b) a different representation of the concept results:

$$|w\rangle = b_0|0_q\rangle + b_1|1_q\rangle, \text{ where } |b_0|^2 + |b_1|^2 = 1.$$

We posit that when concepts or ideas can be described as existing in a superposition state as in (2) and (3), they are experienced consciously as vague or 'half-baked', and indeed experimental evidence for such states has been obtained (Gabora & Saab, 2011). By looking at ideas from these different perspectives, humans can achieve a well-rounded understanding of a concept that is far more detailed than can be achieved in standard formal representations. Indeed, humans frequently encounter situations where looking at a concept from one perspective brings to mind another perspective, and so on, until a very detailed (and frequently quite creative) understanding of the concept has been achieved. Eventually however, a particular interpretation must be settled upon, and this process is described in this formalism through a process of 'measurement', which invokes an associated collapse of the state. Here, the probability that a person will ascribe a particular interpretation to a given concept is proportional to the length of the vector in that dimension (i.e. a projection onto the relevant basis state). This is represented in the formalism by taking the square of the length of the vector along the relevant axis (Isham, 1995; Bruza *et. al*, 2009; Kitto *et. al*, 2011). More formally, the probability that our early human will interpret concept $|w\rangle$ in the sense represented by the basis *p* as $P = |a_1|^2$ which is noticeably different from the interpretation that will be provided to the same concept in context *q* ($P = |b_1|^2$).

We can immediately see that a different context thus results in a different probability value through reference to Fig. 3(a) and (b). In this formalism a different context can result in a very different interpretation becoming likely. Note that we can also extract the probability that a person will not associate a particular interpretation with a given concept ($P = |a_0|^2$ for context *p*, and $P = |b_0|^2$ for context *q*).

Thus, returning to the idea of a FIRE, the probability that it will be interpreted as *dangerous* will be greater in the second context *q* than in the first. Perhaps this might be used to represent the likely danger that an early human will attribute to the concept of FIRE in winter and summer respectively. This allows for a dense representation of the concept FIRE. We do not need to encode each of the different meanings explicitly, they come instead from an interpretation associated to the idea that is attributed at the point of interpretation.

**Second Transition: The 'Big Bang' of Human Creativity**

The European archaeological record indicates that a truly unparalleled cultural transition occurred between 60,000 and 30,000 years ago, at the onset of the Upper Paleolithic (Bar-Yosef, 1994; Klein, 1989; Mellars, 1973, 1989a, 1989b; Soffer, 1994; Stringer & Gamble, 1993). Considering it "evidence of the modern human mind at work," Richard Leakey (1984:93-94) describes the Upper Palaeolithic as "unlike previous eras, when stasis dominated, ... [with] change being measured in millennia rather than hundreds of millennia." Similarly, Mithen (1996) refers to the Upper Paleaolithic as the 'big bang' of human culture, exhibiting more innovation than in the previous six million years of human evolution. This period exhibits the more or less simultaneous appearance of traits considered diagnostic of behavioral modernity. It marks the beginning of a more organized, strategic, season-specific style of hunting involving specific animals at specific sites, elaborate burial sites indicative of ritual and religion, evidence of dance, magic, and totemism, the colonization of Australia, and replacement of Levallois tool technology by blade cores in the Near East. In Europe, complex hearths and many forms of art appeared, including cave paintings of animals, decorated tools and pottery, bone and antler tools with engraved designs, ivory statues of animals and sea shells, and personal decoration such as beads, pendants, and perforated animal teeth, many of which may have indicated social status (White, 1989a, 1989b).

Whether this period was a genuine revolution culminating in behavioral modernity is hotly debated because claims to this effect are based on the European Palaeolithic record, and largely exclude the African record (McBrearty & Brooks, 2000; Henshilwood & Marean, 2003). Indeed, most of the artifacts associated with a rapid transition to behavioral modernity at 40–50,000 years ago in Europe are found in the African Middle Stone Age tens of thousands of years earlier. However the traditional and currently dominant view is that modern behavior appeared in Africa between 40,000 and 50,000 years ago due to biologically evolved cognitive advantages, and spread, replacing existing species, including the Neanderthals in Europe (e.g., Ambrose, 1998; Gamble, 1994; Klein, 2003; Stringer & Gamble, 1993). Thus from this point onward there was only one hominid species: modern *Homo sapien, and* despite lack of overall increase in cranial capacity, their prefrontal cortex, and more particularly their orbitofrontal region, increased significantly in size (Deacon, 1997; Dunbar, 1993; Jerison, 1973; Krasnegor, Lyon, and Goldman-Rakic, 1997; Rumbaugh, 1997) in what was most likely a time of major neural reorganization (Klein, 1999; Henshilwood, d'Errico, Vanhaeren, van Niekerk, and Jacobs, 2000; Pinker, 2002).

Given that the Middle/Upper Palaeolithic was a period of unprecedented creativity, what kind of cognitive processes were involved?

## A Testable Hypothesis

Converging evidence suggests that creativity involves the capacity to shift between two forms of thought (Finke, Ward, & Smith, 1992; Gabora, 2003; Howard-Jones & Murray, 2003; Martindale, 1995; Smith, Ward, & Finke, 1995; Ward, Smith, & Finke, 1999): (1) *Divergent* or *associative* processes are hypothesized to occur during idea generation, while; (2) *convergent* or *analytic* processes predominate during the refinement, implementation, and testing of an idea. It has been proposed that the Paleolithic transition reflects a mutation to the genes involved in the fine-tuning of the biochemical mechanisms underlying the capacity to subconsciously shift between these modes, depending on the situation, by varying the specificity of the activated cognitive receptive field. This is referred to as *contextual focus*[4] because it requires the ability to focus or defocus attention in response to the context or situation one is in. Defocused attention, by diffusely activating a broad region of memory, is conducive to divergent thought; it enables obscure (but potentially relevant) aspects of the situation to come into play. Focused attention is conducive to convergent thought; memory activation is constrained enough to hone in and perform logical mental operations on the most clearly relevant aspects. The theory is consistent with the notion that creativity involves both freedom and constraint; the generation of cultural novelty often starts with structural rules and frameworks (as in the templates of a sonnet or a tragedy) as a basis to deviate from.

[4] In neural net terms, contextual focus amounts to the capacity to spontaneously and subconsciously vary the shape of the activation function, flat for divergent thought and spiky for analytical.

**Support from the Computational Model**

Again, because it would be difficult to empirically determine whether Paleolithic humans became capable of contextual focus, we began by determining whether the hypothesis is at least computationally feasible. To do so, we used an evolutionary art system that generated progressively evolving sequences of artistic portraits. In this context, we sought to determine whether incorporating contextual focus into the fitness function would enable the computer system to generate art that humans find aesthetically pleasing and "creative" on its own (*i.e.*, requiring no human intervention once initiated).

We implemented contextual focus in the evolutionary art algorithm by giving the program the capacity to vary its level of fluidity and control over different phases of the creative process in response to the output it generated. The creative domain of portrait painting was chosen because it requires both focused attention and analytical thought to accomplish the primary goal of creating a resemblance to the portrait sitter, as well as defocused attention and associative thought to deviate from resemblance in a way that is uniquely interesting, *i.e*., to meet the broad and often conflicting criteria of aesthetic art. Since the advent of photography (and earlier), portrait painting has not just been about accurate reproduction, but also about achieving a creative or stylized representation of the sitter. Since judging creative art is subjective, a representative subset of the automatically produced artwork from this system was selected, output to high quality framed images, and submitted to peer reviewed and commissioned art shows, thereby allowing it to be judged positively or negatively as creative by human art curators, reviewers and the art gallery going public.

The software incorporates several techniques that enable it to shift between different modes of thought, which are summarized here. (Implementation details are provided elsewhere (DiPaola, 2009; DiPaola & Gabora, 2007, 2009; Gabora & DiPaola, submitted). Our goal was to build a fitness function incorporating the notion of contextual focus so that the software could shift between small ordered steps and large leaps through the landscape of artistic possibilities. This was carried out as follows: The system's default processing mode is an analytic mode, in which the primary aim is to achieve an accurate resemblance (similarity to the sitter image). Certain functional triggers (such as if the system is 'stuck' and not improving) shift it to a more associative processing mode. This mode aims to achieve painterly aesthetic flair using principles of art creation (rules of composition, tonality, and color theory) as well as portrait knowledge space. Specifically, it takes into account: (1) face versus background composition, (2) tonal similarity over exact color similarity, matched with a sophisticated artistic color space model that weighs for warm-cool color temperature relationships based on analogous and complementary color harmony rules, and (3) unequal dominant and subdominant tone and color rules, and other artistic rules based on a portrait painter knowledge domain as detailed in DiPaola (2009).

Incorporating contextual focus into the computer program not only improved its ability to generate a good resemblance, but resulted in more abstract, aesthetically appealing portraits as well. Humans rated the portraits produced by this version of the portrait painting program with contextual focus as much more creative and interesting than a previous version that did not use contextual focus and, unlike its predecessor, the output of this program generated public attention worldwide. As shown in Figure 4, sample pieces were exhibited at peer reviewed, juried, or commissioned shows in several major galleries and museums that typically only accept human art work, including the TenderPixel Gallery in London, Emily Carr Galley in Vancouver, Kings Art Centre at Cambridge University, the MIT Museum, and the High Museum in Atlanta. The work was also selected for its aesthetic value to accompany a piece in *Nature* (Padian, 2008). While these are subjective measures, they are standard in the art world. Thus using contextual focus the computer program automatically produced novel creative artifacts, both as single art pieces, and as gallery collections of related art with interrelated creative themes, which provides compelling evidence of the effectiveness of contextual focus.

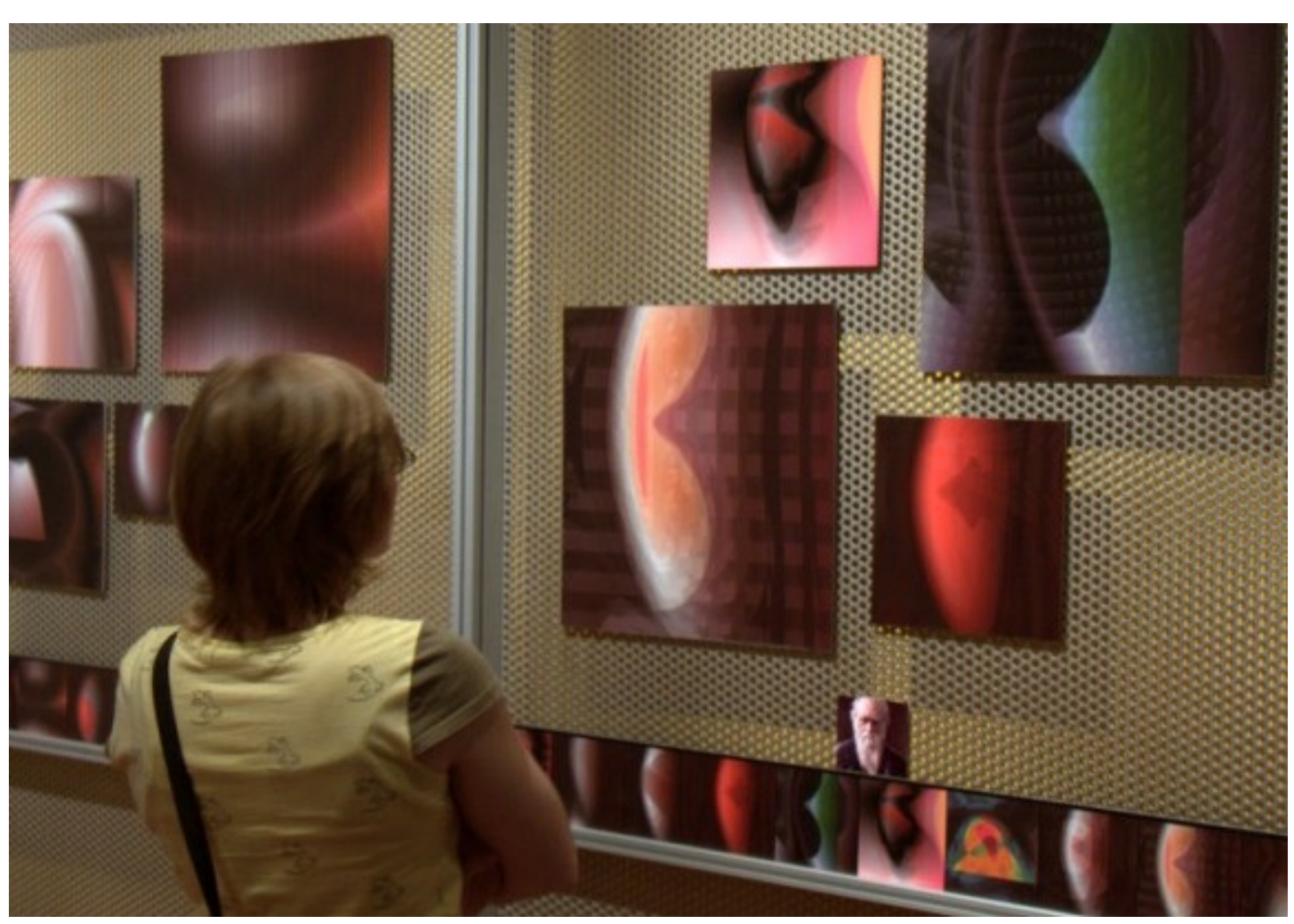

Figure 4. Images produced by a computational art program that uses contextual focus at the MIT Museum in Cambridge, MA. These works have been seen by tens of thousands, and perceived as creative art works on their own by the art public.

In sum, these results support the hypothesis that the impact of recursive recall was vastly accentuated by the capacity to shift between associative and analytic processing modes. This opened up a much greater variety of ways of seeing concepts from different contexts and examining ideas from different perspectives until one converges on an understanding that takes multiple facets into account. We suggest that a mechanism akin to contextual focus is what makes possible the cumulative creativity exhibited by successful computational models of language evolution (*e.g*., Kirby, 2001).

**Modeling Contextual Focus: The Shifting between Convergent and Divergent Thought**

An even more compelling approach would result from developing a cognitive system that is capable of shifting between processing few features or properties of concepts and ideas (analytic or convergent thinking) to encoding many features or properties of concepts and ideas (associative or divergent thinking). The divergent mode would be highly conducive to the emergence of new concept combinations; since there are more properties encoded per concept, there are more potential connections, while the convergent mode would allow for focus and the honing in on useful ideas. Divergent thought is conducive to putting concepts together in new combinations. Using the quantum formalism discussed above, concept combination has been modeled using a tensor product, and other more complex but accurate mathematical structures.

The details of this and related models have been discussed elsewhere (Aerts & Gabora, 2005a,b; Gabora & Aerts, 2002, 2009; Bruza *et. al*, 2009; Kitto *et. al*, 2011). However, the basic idea can be illustrated through a consideration of the two concepts FIRE and FOOD, and how they might have been combined in a creative manner by an early human. These two concepts are likely to have been thought of in a mutually exclusive manner by early humans, as fire would burn forests and fields so decreasing the expected yield of food. Thus, an increased experience of FIRE might have been expected to decrease the yield of food. However, at some point, FIRE was recognized as a tool; it could be used to create more food, by making inedible materials edible, rather than just being recognized as something that would decrease yields by burning food sources, etc. Representing FIRE as a superposition of useful ($|1_p\rangle$),) and not useful ($|0_p\rangle$)), and FOOD as a superposition of edible ($|1_q\rangle$), and inedible ($|0_q\rangle$), we can write the two combined concepts as

$$|FIRE\rangle \otimes |FOOD\rangle$$

$$= \left(a_0|0_p\rangle + a_1|1_p\rangle\right) \otimes \left(x_0|0_q\rangle + x_1|1_q\rangle\right)$$

$$= a_0x_0|0_p\rangle \otimes |0_q\rangle + a_0x_1|0_p\rangle \otimes |1_q\rangle + a_1x_0|1_q\rangle \otimes |0_p\rangle + a_1x_1|1_p\rangle \otimes |1_q\rangle$$

which is a superposition state that arises in the higher dimensional space represented by the four dimensional basis states: $\{|0_p\rangle \otimes |0_q\rangle, |0_p\rangle \otimes |1_q\rangle, |1_p\rangle \otimes |0_q\rangle, |1_p\rangle \otimes |1_q\rangle\}$ (see Isham (1995) for more details about this higher dimensional space).We immediately see that the combination of these two concepts has led to a combinatorial explosion of possibilities; in other words this is a divergent process. If a person is now exposed to another concept, we can imagine a situation where their current cognitive state expands further still, into a yet higher dimensional space. This process might go on for a number of steps, however, this increasingly more complex state is likely to be very difficult to maintain. Indeed, a potential downfall of processing in an associative mode and coming up with unusual combinations is that since effort is devoted to the re-processing of previously acquired material, less effort may be devoted to being on the lookout for danger and simply carrying out practical tasks. Thus, associative thought was of little use until one could have a way of shifting back to a more analytical mode of thought. By re-processing the new combination from increasingly constrained contexts or points of view, it would become clearer how to manifest it. Thus, while some associative thought is indisputably useful, it carries a high cognitive load, which increases as more and more concepts are combined. Eventually, there will be an adaptive advantage in settling upon one particular interpretation in a process of convergence.

The process of 'measurement' discussed above performs this function, even in this scenario of rapidly expanding possibilities, and results in a convergent situation where one idea is finally settled upon, in turn lessening the load associated with maintaining a cognitive state. In the case above, an early human might have realized that FIRE when combined with FOOD could usefully render the inedible edible (as is represented by the state $a_1x_1|1_p\rangle \otimes |1_q\rangle$). The probabilities arising in this scenario might be very small, as the coefficients of equation (5) become smaller with each combination, thus indicating a situation where it is becoming more and more cognitively difficult to settle upon a particular meaning, but also more likely that a highly improbable interpretation might be settled upon. Eventually, if enough humans experience this unusual cognitive state, then there is a significant probability that one of them will start to cook inedible plants, so rendering them edible. Initially it may not be obvious how a new concept combination could make sense or materialize given the constraints of the world it is 'born' into;

for example, one does not know which features of each parent concept are inherited in the combination. However, current work is being directed at finding natural representations of concepts, utilized naturally during the process of combination.

In summary, if early humans reached a stage where they could employ a divergent process of concept combination that was followed by a shift to a more constrained or convergent processing mode, then they would have found themselves at a significant adaptive advantage. They would have been capable of not just creativity, but also directed thought, and so would have reached a new stage of cognitive activity.

**Conclusions and Future Directions**

Since concepts are the building blocks of human cognition, the explanation of how flexible, open-ended cognitive processes of thought arose will require a theory of concepts that can account for and model their contextual, noncompositional behavior. We showed how a quantum-inspired theory of concepts can be used to rigorously flesh out theories concerning the origins of modern cognition. Many species engage in acts that could be said to be creative, but humans are unique in that our creative ideas build on each other cumulatively. Indeed, it is for this reason that culture is widely construed as an evolutionary process (Bentley, Ormerod, & Batty, 2011; Cavalli-Sforza & Feldman, 1981; Gabora, 1996, 2008; Hartley, 2009; Mesoudi, Whiten & Laland, 2004, 2006; Whiten, Hinde, Laland, & Stringer, 2011). Our unique cognitive capacities are revealed in all walks of life, and have transformed the way we live and the planet we live on. We discussed two transitions in the evolution of human cognition: (1) its origins approximately two million years ago, and (2) what has been referred to as the cultural explosion or 'big bang' of human creativity approximately 50,000 years ago. We discussed cognitive mechanisms that have been proposed to underlie these transitions, and summarized efforts to model them, both computationally and mathematically.

It has been hypothesized that the origins of complex human cognition can be attributed to the onset of *recursive recall,* in which one thought or stimulus evokes another in a string of associations (Donald, 1991). This allowed for the chaining together of real or imagined episodes into a stream of thought, or the chaining of movements into complex actions, such that feedback about one component affected performance of the next. This hypothesis has been shown to be compatible with likely changes in the architecture of human memory associated with the increase in cranial capacity at this time (Gabora, 2003, 2008a). Moreover, in a test of this hypothesis using a computational model of cultural evolution in which neural network based agents evolve ideas for actions through invention and imitation, chaining was shown to result in greater cultural diversity, open-ended generation of novelty, no ceiling on the mean fitness of cultural variants, and greater ability to make use of learning (Gabora & Saberi, 2011). This shows that the hypothesis that recursive recall played an important role in the origins of complex cognition is computationally feasible. However, in the computational model we simply compared runs in which agents were limited to single-step actions to runs in which they could chain simple actions into complex ones; chaining did not arise naturally through associative recall due to how items were encoded in memory. We suggest that it is not mere chaining that paved the way for complex cognition and cultural evolution, but chaining that involves the restructuring of concepts by viewing them from different contexts, and proposed that a formal model of this process will be required. We showed how a quantum-inspired theory of concepts can be used to model the transition to a state in which concepts and ideas are encoded in enough detail that associations amongst them are rich enough for a natural chaining through associative recall to occur, resulting in the capacity to progressively shape concepts, ideas, and actions by observing them from different contexts.

We discussed the hypothesis that the explosion of creativity in the Middle/Upper Paleolithic was due to onset of *contextual focus:* the capacity to shift between associative, conducive to forging new concept combinations, and analytic thought, conducive to manifesting them. Incorporating contextual focus (the capacity to shift between analytic and associative modes) into a computational model of portrait painting has resulted in faster convergence on portraits that human observers found more preferable (DiPaola, 2009; DiPaola & Gabora, 2007, 2009). This supports the hypothesis that contextual focus provides a computationally plausible explanation for the cognitive capacities of modern humans.

A limitation of this work was that contextual focus was simply modeled as the capacity to shift between the competing goals of achieving an accurate resemblance of the sitter, and deviating from the sitter's likeness by employing more abstract painterly techniques that exaggerate, minimize, or modify. This chapter also discussed a more sophisticated model of contextual focus using again the quantum-inspired model of concept combination. We show that if a cognitive system is capable of undergoing a transition from encoding few features or properties of concepts and ideas (analytic or convergent thinking) to encoding many features or properties of concepts and ideas (associative or divergent thinking) then new concept combinations are more likely to arise. The drawback is that such associative states are cognitively difficult to maintain, but we showed that if concept combination is followed by a shift to a more constrained or analytic processing mode, then an eventual interpretation can be settled upon, as the new idea or concept emerges from its previously 'half baked' state.

We are currently engaged in the move to more cognitively plausible computational implementations of creativity and its evolution. One of the projects that will soon be underway will implement contextual focus in the EVOC model of cultural evolution that was used for the 'origin of creativity' experiments. This will be carried out as follows. The fitness function will change periodically, so that agents find themselves no longer performing well. They will be able to detect that they are not performing well, and in response, increase the probability of change to any component of a given action. This temporarily makes them more likely to "jump out of a rut" resulting in a very different action, thereby simulating the capacity to shift to a more associative form of thinking. Once their performance starts to improve, the probability of change to any component of a given action will start to decrease to base level, making them less likely to shift to a dramatically different action. This is expected to help them perfect the action they have settled upon, thereby simulating the capacity to shift to a more associative form of thinking.

In short, we have developed several lines of investigation to formally test the feasibility of the hypothesis that human 'mindedness' stems from onset of the capacity to see things in context, or from multiple perspectives. We posit that this began with the onset of representational description at around the time of the appearance of *Homo erectus,* and that it was vastly enhanced by the onset of contextual focus, some time following the appearance of anatomically modern humans. Contextual focus enabled humans to shift between a minimally contextual analytic mode of thought, and a highly contextual associative mode of thought, conducive to 'breaking out of a rut'.

Thus we are on our way toward modeling the mechanisms that could have made possible modern human cognition, along with the subsequent transformation of the planet we live on.

## Acknowledgements

This project was supported in part by the Australian Research Council Discovery grant DP1094974, the Social Sciences and Humanities Research Council of Canada, the Fund for Scientific Research of Flanders, Belgium.